\documentclass[11pt]{revtex4}

\usepackage{hyperref}
\hypersetup{dvips,dvipdfm,colorlinks=true,urlcolor=magenta,filecolor=magenta,linktoc=page,citecolor=red,linkcolor=blue,bookmarks=true}
\usepackage{graphicx}
\usepackage{dcolumn}
\usepackage{color}
\usepackage{bm}
\usepackage{amsmath}
\usepackage{amsfonts}
\usepackage{amssymb}
\usepackage{array}
\usepackage{makecell}
\usepackage[thinlines]{easytable}
\usepackage{multirow}
\usepackage{nicefrac}

\newcounter{defcounter}
\newcolumntype{L}[1]{>{\raggedright\let\newline\\\arraybackslash\hspace{0pt}}m{#1}}
\newcolumntype{C}[1]{>{\centering\let\newline\\\arraybackslash\hspace{0pt}}m{#1}}
\newcolumntype{R}[1]{>{\raggedleft\let\newline\\\arraybackslash\hspace{0pt}}m{#1}}

\input epsf

\begin{document}

\title{Emergent Scenario in first and second order non-equilibrium thermodynamics and stability analysis}

\author{Sourav Haldar\footnote {sourav.math.ju@gmail.com}}
\affiliation{Department of Mathematics, Jadavpur University, Kolkata-700032, West Bengal, India.}

\author{Pritikana Bhandari\footnote {pritikanab@gmail.com}}
\affiliation{Department of Mathematics, Jadavpur University, Kolkata-700032, West Bengal, India.}

\author{Subenoy Chakraborty\footnote {schakraborty.math@gmail.com}}
\affiliation{Department of Mathematics, Jadavpur University, Kolkata-700032, West Bengal, India.}


\begin{abstract}

First and second order non-equilibrium thermodynamics are studied in the context of particle creation mechanism for homogeneous
and isotropic FLRW model and a general formulation of the emergent scenario is investigated. Finally, the stability of the
non-equilibrium thermodynamics is examined.\\\\
{\bf Keywords\,:} Non-equilibrium thermodynamics, Emergent scenario, Particle creation, Bulk viscous pressure.\\\\
PACS Numbers\,: 95.30.Sf, 98.80.Cq, 11.90.+t, 05.70.Ce\\\\
\end{abstract}

\maketitle


\section{Introduction}

Non-equilibrium thermodynamics seems to play a vital role for the description of the Universe in the early eras. Schr\"{o}dinger
\cite{Schrodinger1} initiated such study in the context of the microscopic description of the gravitationally induced particle
production in an expanding universe. From the point of view of quantum field theory in a curved space-time, Parker and others
\cite{Parker1, Birrel1, Mukhanov1} reconsidered this issue. From thermodynamic viewpoint, the study of non-equilibrium thermodynamics
can be classified as first order formalism and second order deviation from equilibrium era. The first order theory due to Eckart
\cite{Eckart1} and Landau and Lifshitz \cite{Landau1} has drawback in relation to causality and stability. These problems are
eliminated in the second order theory due to Muller \cite{Muller1}, Israel \cite{Israel1}, Israel and Stewart \cite{Israel2, Israel3}
and Pavon {\it et al} \cite{Pavon1}. According to them, dissipative phenomena in the form of bulk and shear viscous pressure and heat
flux are considered as dynamical variables with causal evolution equations having subluminal speed of viscous perturbations.

However, in cosmological context, as space-time is usually assumed to be homogeneous and isotropic, so dissipative phenomenon is only
in the form of bulk viscous pressure which may appear either due to interacting cosmic fluids \cite{Weinberg1, Straumann1, Schweizer1, Udey1, Zimdahl1}
 or due to non-conservation of (quantum) particle number \cite{Zeldovich1, Murphy1, Hu1}, which will be considered
in the present work with adiabatic particle production \cite{Prigogine1, Calvao1}, having constant entropy per particle. As a result,
entropy production is carried out by the enlargement of the phase space of the system and particle production rate is linearly related
to the bulk viscous pressure.

The concept of emergent scenario is a very fascinating issue in standard cosmology to look for singularity free inflationary models.
This model of Universe has no time-like singularity, ever existing and having almost static behaviour in the infinite past
$(t \rightarrow -\infty)$ . Asymptotically, this model is familiar to Einstein Static Universe which is a very specific model of a closed Universe which is sourced by positive spatial curvature, dust and a positive cosmological constant. But the scale factor is unstable towards increasing or decreasing from an initial static configuration. Also it can be considered as a modern version and extension of the Lemaitre-Eddington universe.
Historically long back in 1967, Harrison \cite{Harrison1} first proposed a closed model of the Universe which contains different matter sources (than Einstein Static Universe) and which can be stable at the level of the scale factor. Asymptotically, these models are certainly kindred to the Einstein static Universe, but are not quite the same. Subsequently, several authors \cite{Ellis1, Ellis2, Mukherjee1, Mukherjee2, Mulryne1, Banerjee1, Nunes1, Lidsey1, Debnath1, Paul1, Debnath2, Mukerji1, Labrana1, Chakraborty1, Paul2, Bhattacharya1} have formulated this model in different gravity theories and also for different cosmic substratums. Further, as the spatial slices are assumed to be flat in the current thermodynamic model, so it should not correspond to Einstein Static Universe in the asymptotic past.

In thermodynamical context, it is found that consideration of entropy \cite{Gibbons1, Gibbons2} favours the initial state of our
Universe to be the Einstein static state. Also Pavon {\it et al} \cite{delcampo1, Pavon2} has examined the validity of the generalized
second law of thermodynamics in the transition epochs, both from a generic initial Einstein static phase to the inflationary
phase and from inflationary era to the conventional thermal radiation dominated era. In reference \cite{Chakraborty1}, emergent scenario
in the frame work of particle creation process has been studied both for first and second order variation from equilibrium thermodynamics
in absence of chemical potential or specific choice of bulk viscous pressure. The present work is an extension of it. Here, attempts are
made for obtaining emergent scenario for non-zero chemical potential (in 1st order theory) and for a general solution of the bulk
viscous evolution equation (2nd order theory). The paper is organized as follows\,:  Section II deals with basic concepts related to
non-equilibrium thermodynamics in the context of particle creation mechanism. Validity of emergent scenario in the first order
formalism is discussed in section III while section IV deals with the same in the context of second order formulation. Stability criteria due to thermodynamic parameters are investigated in section V. Finally,
summary of the work and possible conclusions are presented in section VI.

\section{Non-equilibrium Thermodynamics in the context of particle creation mechanism}

${}$
\vspace{-0.5cm}\\
In an open thermodynamical system, the number of fluid particles are not conserved $\left(N_{~;\mu}^\mu \neq 0 \right)$ due to particle creation
and it is described as \cite{Chakraborty1, Chakraborty2, Zimdahl2}
\begin{equation} \label{eq1}
\dot{n} + \theta n = n \Gamma \,.
\end{equation}
Here the particle flow vector $(N^\mu)$ is related to the particle four velocity $(u^\mu)$ as $N^\mu = nu^\mu$ , $\theta = u_{~;\mu}^\mu$
is the fluid expansion, $\Gamma$ is the rate of change of the particle number in a co-moving volume $V$ , $n = \frac{N}{V}$ is the number
density with $N$\,, the present number of particles in $V$ and notationally, $\dot{n} = n_{,\mu}u^\mu$ . The Gibb's equation can be written as
\begin{equation} \label{eq2}
T ds = dq = d\left(\frac{\rho}{n}\right) + pd\left(\frac{1}{n}\right)\,,
\end{equation}
where $s$ is the entropy per particle, $dq = \frac{dQ}{N}$ is the amount of heat per particle and $T$ is the temperature of the fluid.
For the present open system, as the particle number $n$ is not conserved so, the above Gibb's equation takes the form
\begin{equation} \label{eq3}
T dS = dQ = d(\rho V) + pdV - \left(\frac{h}{N}\right)d(N) \equiv dh -Vdp - \frac{h}{N}dN
\end{equation}
where $h = (\rho + p)V$ is the enthalpy of the fluid.

The above equation (\ref{eq3}) can be interpreted as the first law of thermodynamics for the open system under consideration. Now, if the
thermodynamical system is assumed to be adiabatic in nature ($i.e.$ there is no heat propagation) then the above modified first law of
thermodynamics can be written as
\begin{equation} \label{eq4}
d\rho + (\rho + p + \Pi)d\ln V = 0
\end{equation}
\begin{equation} \label{eq5}
\mbox{with}~~~~~~~\Pi = -\frac{h}{N}\frac{d(N)}{dV}~,
\end{equation}
an additional pressure due to matter creation. For homogeneous and isotropic FLRW space-time model eq.\,(\ref{eq4}) can be written as
\begin{equation} \label{eq6}
\dot{\rho} + 3H(\rho + p + \Pi) = 0
\end{equation}
which is nothing but the conservation equation $T_{\nu ; \mu}^{\mu} = 0$ for the matter field having energy-momentum tensor $T_{\mu \nu}$ as
\begin{equation} \label{eq7}
T_{\mu \nu} = (\rho + p + \Pi)u_\mu u_\nu + (p + \Pi)g_{\mu \nu}
\end{equation}
with $u^\mu$ , the fluid four velocity. Thus the extra (dissipative) pressure term $\Pi$ due to particle creation mechanism can be interpreted
as bulk viscous pressure of relativistic fluid.

Using (\ref{eq1}) in equation (\ref{eq5}) one gets the dissipative pressure $\Pi$ in terms of the particle creation rate as (for adiabatic process)
\cite{Chakraborty1, Chakraborty2}
\begin{equation} \label{eq8}
\Pi = -\frac{\Gamma}{3H}(\rho + p)\,.
\end{equation}
Also, the Friedmann equations for FLRW model are
\begin{equation} \label{eq9}
3H^2 = \kappa \rho ~~~~\mbox{and}~~~~ 2\dot{H} = -\kappa(\rho + p + \Pi)\,.
\end{equation}
Thus eliminating $\Pi$ between (\ref{eq8}) and (\ref{eq9}) we have
\begin{equation} \label{eq10}
\frac{\Gamma}{3H} = 1 + \frac{2}{3(1+\omega)}\left(\frac{\dot{H}}{H^2}\right)
 \end{equation}
with $\omega$ , the adiabatic index ($i.e.$ $p=\omega \rho$) . So the above equation can be used in both ways --- particle creation rate
can be determined if cosmic evolution can be obtained from the particle creation rate as a known function of the Hubble parameter $H$\,.

\section{First order non-equilibrium thermodynamics and Emergent Scenario}

${}$
\vspace{-0.5cm}\\
In first order theory the entropy flow vector due to Eckart \cite{Eckart1, Chakraborty2, Zimdahl2} is defined as
\begin{equation} \label{eq11}
s_E^\mu = nsu^\mu
\end{equation}
so that one gets
\begin{equation} \label{eq12}
(s_E^\mu)_{;\mu} = -\frac{\Pi}{T}\left(3H + \frac{n\mu \Gamma _E}{\Pi _E}\right)~,
\end{equation}
using the number conservation equation (\ref{eq1}) and the matter conservation equation (\ref{eq6}) . In the above equation (\ref{eq12}) ,
the chemical potential `$\mu$' has the expression
\begin{equation} \label{eq13}
\mu = \left(\frac{\rho + p}{n}\right) - Ts~.
\end{equation}
For the adiabatic process using (\ref{eq8}) in the entropy flow equation (\ref{eq12}) , the variation of the entropy flow can be simplified as
\begin{equation} \label{eq14}
(s_E^\mu)_{;\mu} = ns \Gamma _E~.
\end{equation}
Thus the second law of thermodynamics ($i.e.$ $(s_E^\mu)_{;\mu} \geq 0$) is always satisfied when there is creation of particles ($i.e.$ $\Gamma _E >0$)\,.

In particular, if ~$\Gamma _E = \Gamma _0 + \Gamma _1 H$ (with $\Gamma _0 >0\,,~\Gamma _1 \geq 0$) then from equation (\ref{eq10})\,, the evolution equation takes the form
\begin{equation} \label{eq15}
	\dot{H} = \frac{3(1+\omega)}{2}\left[\frac{\Gamma _0}{3}H + \left(\frac{\Gamma _1}{3}-1\right)H^2\right]~,
\end{equation}
which on integration gives the Hubble parameter as
\begin{equation} \label{eq16}
	H = \frac{d}{\left[\exp (-\mu \tau) + \sigma \right]} 
\end{equation}
where~ $d= \frac{\Gamma _0}{3}\,,~\sigma = 1-\frac{\Gamma _1}{3}\,,~\mu = \frac{\Gamma _0}{2}(1+\omega)\,,~\tau = t - t_0$ ~and ~$t_0$~ is the constant of integration.\\
So the explicit form of the scale factor can be obtained by integrating once more as
\vspace{2mm}

{\bf (\bm{$i$}) \underline{when \bm{$\sigma >0$} ~\bm{$i.e.$}~ \bm{$\Gamma _1 <3$}}}
\begin{equation} \label{eq17}
	\left(\frac{a}{a_0}\right)^\alpha = \left[1 + \sigma \exp (\mu \tau)\right]
\end{equation}

~~~~where ~$\alpha = \frac{3}{2}(1+\omega)\left(1-\frac{\Gamma _1}{3}\right)$ ~and~ $a_0$ ~is the integration constant.\\

{\bf (\bm{$ii$}) \underline{when \bm{$\sigma <0$} ~\bm{$i.e.$}~ \bm{$\Gamma _1 >3$}}} ~~$i.e.$~ $\sigma = -\sigma _0$~,~~$\sigma _0 = \frac{\Gamma _1}{3}-1 > 0$
\begin{equation} \label{eq18}
	\left(\frac{a}{a_0}\right)^{-\alpha _0} = -1 + \sigma _0 \exp (\mu \tau)~~~,~~~~\alpha _0 = \frac{3}{2}(1+\omega) \sigma _0~.
\end{equation}
The above solutions have the following asymptotic or limiting features \cite{Chakraborty1} :
\vspace{1mm}

{\bf (\bm{$i$}) : ~~\bm{$\sigma >0$} :}

~~~~~~~(a)~~$a \rightarrow a_0~,~H \rightarrow 0~~\mbox{as}~~t \rightarrow -\infty$

~~~~~~~(b)~~$a \simeq a_0~,~H \simeq 0~~\mbox{for}~~t \ll t_0$

~~~~~~~(c)~~$a \simeq \exp\left\{\frac{\Gamma _0}{3\sigma}(t-t_0)\right\}~,~H \simeq \left(\frac{d}{\sigma}\right) ~~\mbox{for}~~t \gg t_0$ .
\vspace{1mm}

{\bf (\bm{$ii$}) : ~~\bm{$\sigma = -\sigma _0 < 0$} :}

~~~~~~~~~There is big-rip singularity at~ $t=t_0 + \frac{1}{\mu}\ln \left(\frac{1}{\sigma _0}\right)$ .\\

Therefore, in the first order Eckart theory \cite{Eckart1} the particle production process gives rise to a cosmological solution $(\sigma > 0)$ that describes a scenario of emergent universe.

In this context, one may note that emergent scenario in particle creation context is not new in the literature. In recent past, it has been shown \cite{Chakraborty1} that emergent scenario is possible for both first and second order formulation of non-equilibrium thermodynamics. In that work, the dissipative bulk viscous pressure is chosen in a typical form and chemical potential is assumed to be zero (for the first order theory). Recently, Ghosh and Gangopadhyay \cite{Ghosh1} have shown emergent scenario with non-zero chemical potential but they also assume a specific choice for the dissipative (bulk viscous) pressure. The present work is a more general one in the sense that neither the chemical potential is restricted nor there is any choice of the bulk viscous pressure. Further, in both the previous works (in ref.\,\cite{Chakraborty1} and \cite{Ghosh1}) the particle creation rate is assumed to be constant (the constancy of $\Gamma$ can be seen from eq.\,(35) using Einstein's field equations in ref.\,\cite{Ghosh1}) but in the present work the particle creation rate is chosen as a linear function of $H$\,. Also in the present work, it is found that, if $0 \leq \Gamma _1 <3$ then singularity-free emergent cosmological solution is obtained while we have the usual big-bang cosmology for $\Gamma _1 > 3$\,.

\section{Non-equilibrium thermodynamics in second order formulation and emergent scenario}

${}$
\vspace{-0.5cm}\\
The entropy flow vector $(S^a)$ due to Israel and Stewart \cite{Israel2, Israel3, Zimdahl2} in the second order formulation of non-equilibrium thermodynamics has the expression
\begin{equation} \label{eq19}
S^a = sN^a - \frac{\tau \Pi ^2}{2\zeta T}u^a
\end{equation}
where $T$ is the fluid temperature, $\zeta$ is the coefficient of the bulk viscosity and $\tau$ is the time of relaxation. From the Gibbs equation (\ref{eq2}) the rate of change of entropy per particle has the expression
\begin{equation} \label{eq20}
\dot{s} = -\frac{1}{nT}\left\{\theta \Pi + (\rho + p)\Gamma\right\}\,.
\end{equation}
Using equations (\ref{eq1}) and (\ref{eq20}) one obtains the variation of the entropy flow vector as
\begin{equation} \label{eq21}
S_{~;a}^a = -\frac{\Pi}{T}\left[\theta + \frac{\tau \dot{\Pi}}{\zeta} + \frac{\Pi \tau}{2\zeta}\left\{\theta + \frac{\dot{\tau}}{\tau} - \frac{\dot{\zeta}}{\zeta} - \frac{\dot{T}}{T} \right\}\right] - \frac{n\Gamma}{T}\mu \,.
\end{equation}

For isentropic thermodynamical system the entropy per particle is constant and as a result, (\ref{eq21}) simplifies to
\begin{equation} \label{eq22}
S_{~;a}^a = -\frac{\Pi}{T}\left[\frac{\tau \dot{\Pi}}{\zeta} + \frac{\Pi \tau}{2\zeta}\left\{\theta + \frac{\dot{\tau}}{\tau} - \frac{\dot{\zeta}}{\zeta} - \frac{\dot{T}}{T} \right\} + \frac{n\theta T s}{\rho + p}\right]\,.
\end{equation}
Now choosing the general ansatz as
\begin{equation} \label{eq23}
\tau \dot{\Pi} + \frac{\Pi \tau}{2}\left\{\theta + \frac{\dot{\tau}}{\tau} - \frac{\dot{\zeta}}{\zeta} - \frac{\dot{T}}{T} \right\} + \frac{\zeta n\theta T s}{\rho(1+\omega)} = -\Pi
\end{equation}
one gets
$$S_{~;a}^a = \frac{\Pi ^2}{T\zeta} \geq 0\,.$$
Thus second law of thermodynamics holds and dissipative pressure $\Pi$ has the evolution equation
\begin{equation} \label{eq24}
\frac{\dot{\Pi}}{\Pi} + \frac{1}{2}\left(\theta + \frac{\dot{\tau}}{\tau} - \frac{\dot{\zeta}}{\zeta} - \frac{\dot{T}}{T} \right) + \frac{1}{\tau}\left\{1 + \frac{\zeta n\theta T s}{\rho(1+\omega)\Pi}\right\} = 0\,.
\end{equation}
If we choose the linear relation between dissipative pressure $\Pi$ and the chemical potential $\mu$ as
\begin{equation} \label{eq25}
\Pi = -\zeta \theta \left(1-\frac{\mu n}{\rho + p}\right)
\end{equation}
then the above evolution equation has the simple solution
\begin{equation} \label{eq26}
\Pi ^2 = \Pi _0 ^2 \frac{\zeta T}{a^3 \tau}\,.
\end{equation}
Hence the chemical potential has the explicit form in terms of relaxation time as
\begin{equation} \label{eq27}
\mu = \frac{(\rho + p)}{n}\left[1 + \frac{\Pi _0 T^{\nicefrac{1}{2}}}{\sqrt{\zeta}\theta a^{\nicefrac{3}{2}} \tau ^{\nicefrac{1}{2}}}\right]\,.
\end{equation}
For isentropic thermodynamical process the particle creation rate has the explicit expression
\begin{equation} \label{eq28}
\Gamma = -\frac{\kappa \Pi _0}{(1+\omega)H}\sqrt{\frac{\zeta T}{a^3 \tau}}\,.
\end{equation}
Also for this adiabatic process, the evolution of the thermodynamical parameters can be described as \cite{Chakraborty2, Zimdahl2}
\begin{eqnarray} \label{eq29}
\left. \begin{array}{l}
\dot{\rho} = -(\theta - \Gamma)(\rho + p)~~~,~~~~\dot{p} = -c_s^2(\theta - \Gamma)(\rho + p)\\
\frac{\dot{n}}{n} = -(\theta - \Gamma)~~~,~~~~\frac{\dot{T}}{T} = -(\theta - \Gamma)\frac{\partial p}{\partial \rho}
\end{array} \right.
\end{eqnarray}
where ~$c_s^2 = \frac{\partial p}{\partial \rho}$~ is the adiabatic sound speed. However, in a viscous medium, the sound speed $c_{vs} (<1)$ is given by \cite{Zimdahl3, Hiscock1}
\begin{equation} \label{eq30}
c_{vs}^2 = c_s^2 + c_d^2
\end{equation}
where $c_d^2 = \frac{\zeta}{(\rho + p)\tau}$ , is the velocity of propagation of the dissipative pulses. Further, due to causality, we have $c_d^2 \leq 1 - c_s^2$\,.

Moreover, using the evolution equations (\ref{eq29}) for the thermodynamical parameters and the velocity $c_d$ for the dissipative pulses, the evolution equation (\ref{eq24}) for the dissipative pressure $\Pi$ can also be written in the form
\begin{equation} \label{eq31}
\frac{\dot{\Pi}}{\Pi} + \frac{1}{2}\left\{2(1+\omega)\theta - (1+2\omega)\Gamma - 2\frac{\dot{c_d}}{c_d}\right\} + \left(\frac{1}{\tau} + \frac{n\theta T s c_d^2}{\Pi}\right) = 0\,.
\end{equation}

Now similar to first order Eckart theory, if the particle creation rate is chosen linearly to Hubble parameter $i.e.$ $\Gamma = \Gamma _0 + \Gamma _1 H$\,, then the solution for the bulk viscous pressure ($i.e.$ eq.\,(\ref{eq26})) can be written as
\begin{equation} \label{eq32}
\Pi = -\frac{(1+\omega)}{\kappa}H\left(\Gamma _0 + \Gamma _1 H\right)
\end{equation}
and the bulk viscous coefficient, cosmic temperature and relaxation time are interrelated as
$$\frac{\zeta T}{a^3 \tau} = \frac{(1+\omega) ^2}{\kappa ^2 \Pi _0^2}H^2\left(\Gamma _0 + \Gamma _1 H\right)^2\,.$$
Also from the evolution equation (\ref{eq10}) we have the same emergent solution (for $0<\Gamma _1 <3$) as before. It should be noted that using $\Pi$ from eq.\,(\ref{eq32}), equation (\ref{eq31}) can be considered as the evolution equation for the velocity propagation of the dissipative pulses.

To have a comparative study with the earlier works, we mention that the claim in ref.\,\cite{Ghosh1} for the non-existence of the emergent scenario in the second order formalism is not true. In their causal evolution equation (56) if ~$\Pi = -3\zeta H$ ~is substituted then there are two possibilities
$$(i) ~~\tau = 0 ~~~~~~~~~~~\mbox{or}~~~~(ii) ~~\frac{2\dot{\Pi}}{\Pi} = -\left(3H + \frac{\dot{\tau}}{\tau}-\frac{\dot{\zeta}}{\zeta}-\frac{\dot{T}}{T}\right)~.$$
The authors of ref.\,\cite{Ghosh1} have considered only the first possibility. However, for the second choice, $\Pi$ integrates to
$$\Pi ^2 = \frac{\zeta T}{a^3 \tau}\,.$$
Using ~$\Pi = -3\zeta H$ , one gets a relation
$$a^3 \tau \zeta= \frac{T}{9H^2}$$
which shows that the thermodynamic parameters are very much related to cosmic evolution. Hence the emergent scenario for the second order Israel and Stewart theory in ref.\,\cite{Chakraborty1} is justified. In the present work, we have a more general result for the non-equilibrium thermodynamics in second order formulation and it is found that emergent scenario is possible for non-constant particle creation rate.

\section{Thermodynamic aspects of Particle creation mechanism and Stability criteria}

In particle creation mechanism, the number of fluid particles are not conserved (see eq.\,(\ref{eq1})) and hence the modified first law of thermodynamics for this open system takes the form of equation (\ref{eq3})\,. However, for reversible adiabatic nature of the system there should not be any heat flow and consequently the above modified first law gives the energy conservation relation
\begin{equation} \label{eq33}
d\ln \rho+(1+\widetilde{\omega})d\ln V=0
\end{equation}
where $V = V_0 a^3(t)$ is the physical volume of the Universe at a given time (the index `0' stands for the present volume with $a_0=1$) and~ $\widetilde{\omega} = \omega + \frac{\Pi}{\rho}$~ is the effective equation of state parameter. Here, the effective dissipative bulk viscous pressure $\Pi$ is related to the enthalpy density by the relation (\ref{eq5})\,. Now, if we consider the temperature ($T$) and volume ($V$) as the basic thermodynamical variables then $\rho = \rho (T,V)$\,, and $dS$ in eq.\,(\ref{eq3}) to be exact differential leads to
\begin{equation} \label{eq34}
d\ln T= d\ln \rho + d\ln V +d\ln \left|1+\widetilde{\omega}\right|\,.
\end{equation}
In driving the above equation, the isentropic condition (\ref{eq8}) is used and $\Gamma = \Gamma _0+\Gamma _1H$ is chosen. Using the conservation equation (\ref{eq33}) one can integrate (\ref{eq34}) to give
\begin{equation} \label{eq35}
\frac{T}{T_0} = \left(\frac{1+\widetilde{\omega}}{1+\widetilde{\omega}_0}\right)\left(\frac{\rho}{\rho _0}\right)a^3
\end{equation}
$$i.e.~~~(1+\widetilde{\omega})\frac{\rho V}{T}= (1+\widetilde{\omega}_0)\frac{\rho _0V_0}{T_0} = \mbox{constant}\,.$$
Thus the ideal gas law is also true in the framework of isentropic particle creation mechanism. Using this ideal gas law, the internal energy of the fluid can be written as
\begin{equation} \label{eq36}
	U=\left(\frac{1+\widetilde{\omega}_0}{1+\widetilde{\omega}}\right)\frac{U_0}{T_0}T\,.
\end{equation}

Further, there are derived thermodynamical parameters, namely heat capacity, compressiblity and thermal expansibility which are not only measurable experimentally for any terrestrial fluid but also they are related to the thermal and mechanical stability. In the following, we shall determine the restriction on the equation of state parameter so that the cosmic scenario at the emergent phase is a stable configuration for both first and second order theory of non-equilibrium thermodynamics.
Now choosing temperature and pressure as the basic thermodynamical variables one can write the volume variation as
\begin{equation} \label{eq37}
	\frac{dV}{V} = \alpha dT - \kappa _T dp
\end{equation}
where the thermal expansivity $\alpha$ and isothermal compressibility $\kappa _T$ are defined as
\begin{equation} \label{eq38}
	\alpha = \frac{1}{V}\left(\frac{\partial V}{\partial T}\right)_p~~~,~~~~\kappa _T = -\frac{1}{V}\left(\frac{\partial V}{\partial p}\right)_T
\end{equation}
so that
\begin{equation} \label{eq39}
	\frac{\alpha}{\kappa _T} = \left(\frac{\partial p}{\partial T}\right)_V~.
\end{equation}
Similarly in case of adiabatic process, one has to consider entropy to be constant (instead of temperature) and analogously there is adiabatic compressibility $\kappa _s$\,, and these two compressibilities are related to the heat capacities by the relation
\begin{equation} \label{eq40}
	\frac{\kappa _s}{\kappa _T} = \frac{C_v}{C_p}
\end{equation}
where $C_v$ and $C_p$ are respectively the fluid's heat capacities at constant volume and constant pressure. For the present dissipative type cosmic fluid the interrelation between heat capacities are given by
\begin{equation} \label{eq41}
	C_v = \frac{d\ln V}{(1+\widetilde{\omega})d\ln V-d\ln \widetilde{\omega}}C_p\,.
\end{equation}
One can also obtain $C_v$ from ideal gas law (\ref{eq36}) as
\begin{equation}\label{eq42}
C_v =\frac{pV}{\widetilde{w}T}\,.
\end{equation} 
Now using the expression for internal energy from (\ref{eq36}) and the integrability condition (\ref{eq34}), one gets,
\begin{equation}\label{eq43}
C_p = (1+\widetilde{\omega})\frac{\rho V}{T}= (1+\widetilde{\omega}_0)\frac{\rho _0 V_0}{T_0}= \mbox{constant}\,.
\end{equation}
Using (\ref{eq34}) in the definition (\ref{eq38}) for $\alpha$ one gets
\begin{equation}\label{eq44}
\alpha=\frac{C_p}{(1+\widetilde{\omega}) \rho V}\left[1+\frac{d\widetilde{\omega}}{\widetilde{\omega} \{d\widetilde{\omega}-\widetilde{\omega} (1+\widetilde{\omega})d\ln V\}}\right]\,.
\end{equation} 
Now from eq.\,(\ref{eq39}) using eq.\,(\ref{eq42}), one obtains
\begin{equation} \label{eq45}
	\kappa _T = \frac{\alpha V}{C_p}
\end{equation}
so that from the relation (\ref{eq40})
\begin{equation} \label{eq46}
	\kappa _s = \frac{\alpha V C_v}{C^2_p}\,.
\end{equation}
The thermal stability of a system needs the second order variation of a system to be positive definite $i.e.$
\begin{equation} \label{eq47}
	\delta ^2U = \delta T\delta S - \delta p\,\delta V > 0\,.
\end{equation}
Now choosing $(T,V)$ or $(S,p)$ as the independent thermodynamical variables one can write the above second order variation as
\begin{eqnarray} \label{eq48}
	\delta ^2U =
	\left. \begin{array}{l}
		\frac{C_v}{T}\delta T^2 + \frac{1}{V\kappa _T}\delta V^2 \\
		\mbox{~~~~~~~~~or} \\
		\frac{T}{C_p}\delta S^2 + V\kappa _s\delta p^2 \,.
	\end{array} \right.
\end{eqnarray}
Hence for stability we have
\begin{equation} \label{eq49}
	C_v~,~C_p~,~\kappa _T~,~\kappa _s \geq 0\,.
\end{equation}

In the following, we have presented in tabular form the stability criteria for different range of the equation of state parameter $\widetilde{\omega}$\,.

\begin{center}
	\begin{table}[ht]
		\renewcommand{\arraystretch}{2.5}
		\caption{Conditions for Stability Criteria} \label{tab:1}
		\begin{tabular}{| >{\centering\arraybackslash}m{4cm}|>{\centering\arraybackslash}m{6.5cm}|>{\centering\arraybackslash}m{5.5cm}|}
			\hline
			{\bf Restriction on} \bm{$\widetilde{\omega}$} & {\bf Constraint on the rate of particle creation} \bm{$\Gamma = \Gamma _0+ \Gamma _1H$} & {\bf Stability Condition}\\
			\hline
			$1+\omega >0$ , $~\widetilde{\omega} >0$ & \thead{$\frac{\Gamma_0}{3-\Gamma _1} < H < \frac{\Gamma_0}{\frac{3\omega}{1+\omega}-\Gamma _1}~,~~\omega < \frac{\Gamma _1}{3-\Gamma _1}$ \vspace{2mm}\\ $\mbox{and}~~~~H>\frac{\Gamma_0}{\frac{3\omega}{1+\omega}-\Gamma _1}~,~~\omega > \frac{\Gamma _1}{3-\Gamma _1}$} & $\frac{d\ln \widetilde{\omega}}{d\ln V}<\widetilde{\omega} ~~\mbox{or}~~  \frac{d\ln \widetilde{\omega}}{d\ln V} > (1+\widetilde{\omega})$ \\
			\hline
			$1+\omega >0$ , ~$\widetilde{\omega} <0$ & \thead{$H>\frac{\Gamma_0}{\frac{3\omega}{1+\omega}-\Gamma _1}~,~\omega < \frac{\Gamma _1}{3-\Gamma _1}~~~~\mbox{and}$ \vspace{2mm}\\ $\frac{\Gamma_0}{3-\Gamma _1} < H < \frac{\Gamma_0}{\frac{3\omega}{1+\omega}-\Gamma _1}~,~\omega > \frac{\Gamma _1}{3-\Gamma _1}$} & $\widetilde{\omega}<\frac{d\ln \widetilde{\omega}}{d\ln V} < (1+\widetilde{\omega})$ \\
			\hline
			$1+\omega <0$ , ~$\widetilde{\omega} >0$ & \thead{$H<\frac{\Gamma_0}{\frac{3\omega}{1+\omega}-\Gamma _1}~,~\omega < \frac{\Gamma _1}{3-\Gamma _1}~~~~\mbox{and}$ \vspace{2mm}\\ $\frac{\Gamma_0}{\frac{3\omega}{1+\omega}-\Gamma _1} < H <\frac{\Gamma_0}{3-\Gamma _1} ~,~\omega > \frac{\Gamma _1}{3-\Gamma _1}$} & $\frac{d\ln \widetilde{\omega}}{d\ln V}<{\widetilde{\omega}} ~~\mbox{or}~~  \frac{d\ln \widetilde{\omega}}{d\ln V} > (1+\widetilde{\omega})$ \\
			\hline
			$1+\omega <0$ , ~$\widetilde{\omega} <0$ & \thead{$\frac{\Gamma_0}{\frac{3\omega}{1+\omega}-\Gamma _1} < H <\frac{\Gamma_0}{3-\Gamma _1} ~,~\omega < \frac{\Gamma _1}{3-\Gamma _1}$ \vspace{2mm}\\ $\mbox{and}~~~~H<\frac{\Gamma_0}{\frac{3\omega}{1+\omega}-\Gamma _1}~,~\omega > \frac{\Gamma _1}{3-\Gamma _1}$} & $\widetilde{\omega}<\frac{d\ln \widetilde{\omega}}{d\ln V} < (1+\widetilde{\omega})$ \\
			\hline
			
		\end{tabular}
	\end{table}
\end{center}

\section{Summary of the work}

The motivation of the present work is two fold --- it is a generalization of our earlier work (ref.\,\cite{Chakraborty1}) on (both first and second order formulation) of non-equilibrium thermodynamics in the perspective of particle creation mechanism. Secondly, we have tried to address the criticisms raised by the authors in ref.\,\cite{Ghosh1} about the earlier work ($i.e.$ ref.\,\cite{Chakraborty1}).

In the present work, it has been shown that emergent scenario is possible for both first and second order theory with non-constant particle creation rate (a linear function of the Hubble parameter). Also the thermodynamical variables are not adhocly chosen, rather they are determined from the thermodynamical as well as evolution equations and the assumption of adiabatic process. The restriction on one of the arbitrary parameters in the linear choice of the particle creation rate is crucial for emergent solution, otherwise one gets the usual big bang cosmology.

The present first order theory is not only a generalization of our earlier work but also the work in ref.\,\cite{Ghosh1} and consequently the issues raised by the authors in ref.\,\cite{Ghosh1} are automatically addressed. For the second order theory we have not only shown emergent scenario with a much more general prescription but also have given justifications for the emergent scenario with second order theory in the earlier work. Therefore, we conclude that emergent scenario is a consequence of the general prescription of the (1st and 2nd order) non-equilibrium thermodynamics with particle creation mechanism and it has the basic features of Steady State theory of Fred Hoyle $et~al$ \cite{Hoyle1, Hoyle2, Hoyle3}\,.

Finally, we like to mention the general question whether the present scenario is consistent with the singularity theorems of Hawking
and Penrose \cite{Hawking1}. According to them it is not possible to robustly ({\it i.e.} outside a set of measure zero) avoid an initial singularity in flat or open FRW space time without violating the null energy condition (which may lead to inconsistencies with quantum mechanics) or violating the strong energy condition in a universe with positive spatial curvature. However in the present work the matter is perfect fluid in nature with variable equation of state and it is not necessary to violate the null energy condition for thermal stability and hence we have emergent scenario, not consistent with singularity theorems.

\section{Acknowledgement}

The author SH acknowledges UGC-JRF fellowship and PB acknowledges DST-INSPIRE for awarding Research fellowship. The author SC is thankful to the Inter-University Centre for Astronomy and Astrophysics\,(IUCAA), Pune, India for research facilities at Library. SC also acknowledges the UGC-DRS Programme in the Department of Mathematics, Jadavpur University.\\\\


\end{document}